\documentclass{ws-procs975x65}

\begin{document}

\title{QUANTUM PHASE TRANSITIONS ON PERCOLATING LATTICES}

\author{THOMAS VOJTA and JOS\'{E}~A. HOYOS}

\address{Department of Physics, University of Missouri-Rolla,\\
Rolla, Missouri 65409, USA\\ E-mail: vojtat@umr.edu}

\begin{abstract}
When a quantum many-particle system exists on a randomly diluted lattice, its intrinsic
thermal and quantum fluctuations coexist with geometric fluctuations due to percolation.
In this paper, we explore how the interplay of these fluctuations influences the phase
transition at the percolation threshold. While it is well known that thermal fluctuations
generically destroy long-range order on the critical percolation cluster, the effects of
quantum fluctuations are more subtle. In diluted quantum magnets with and without
dissipation, this leads to novel universality classes for the zero-temperature
percolation quantum phase transition. Observables involving dynamical correlations
display nonclassical scaling behavior that can nonetheless be determined exactly in two
dimensions.
\end{abstract}



\bodymatter

\section{Introduction}\label{sec:intro}

In disordered quantum many-particle systems, random fluctuations due to impurities and
defects coexist with quantum fluctuations and thermal fluctuations. Close to phase
transitions, the interplay between these different types of fluctuations can cause many
unconventional phenomena such as quantum Griffiths
effects\cite{ThillHuse95,RiegerYoung96}, infinite-randomness critical
points\cite{Fisher92,Fisher95} and the destruction of the phase transition by
smearing\cite{Vojta03a} (for a recent review see, e.g., Ref.~\refcite{Vojta06}). A
particular interesting case of this scenario are randomly diluted magnets. Site or bond
dilution defines a percolation problem\cite{StaufferAharony_book91} for the lattice which
can undergo a \emph{geometric} phase transition between a disconnected and a percolating
phase.

Here, we discuss how the interplay between the quantum fluctuations of
the spins and the geometric fluctuations of the lattice changes the phase diagram and the
nature of the magnetic percolation transition. Our paper is organized as follows:
In Sec.\ \ref{sec:percolation}, we
collect the basic results of classical percolation theory\cite{StaufferAharony_book91} to
the extent necessary for the following sections. Section \ref{sec:classical} is devoted
to the behavior of classical magnets on percolating lattices. These are older results
summarized here mainly for comparison with the quantum case. Section \ref{sec:quantum} is
the main part of the paper. Here, we discuss three different examples of percolation
quantum phase transitions (QPTs) of diluted quantum magnets. We conclude in Sec.\
\ref{sec:conclusions}.

\section{Geometric Percolation}\label{sec:percolation}

\begin{figure}[t]
\centerline{\psfig{file=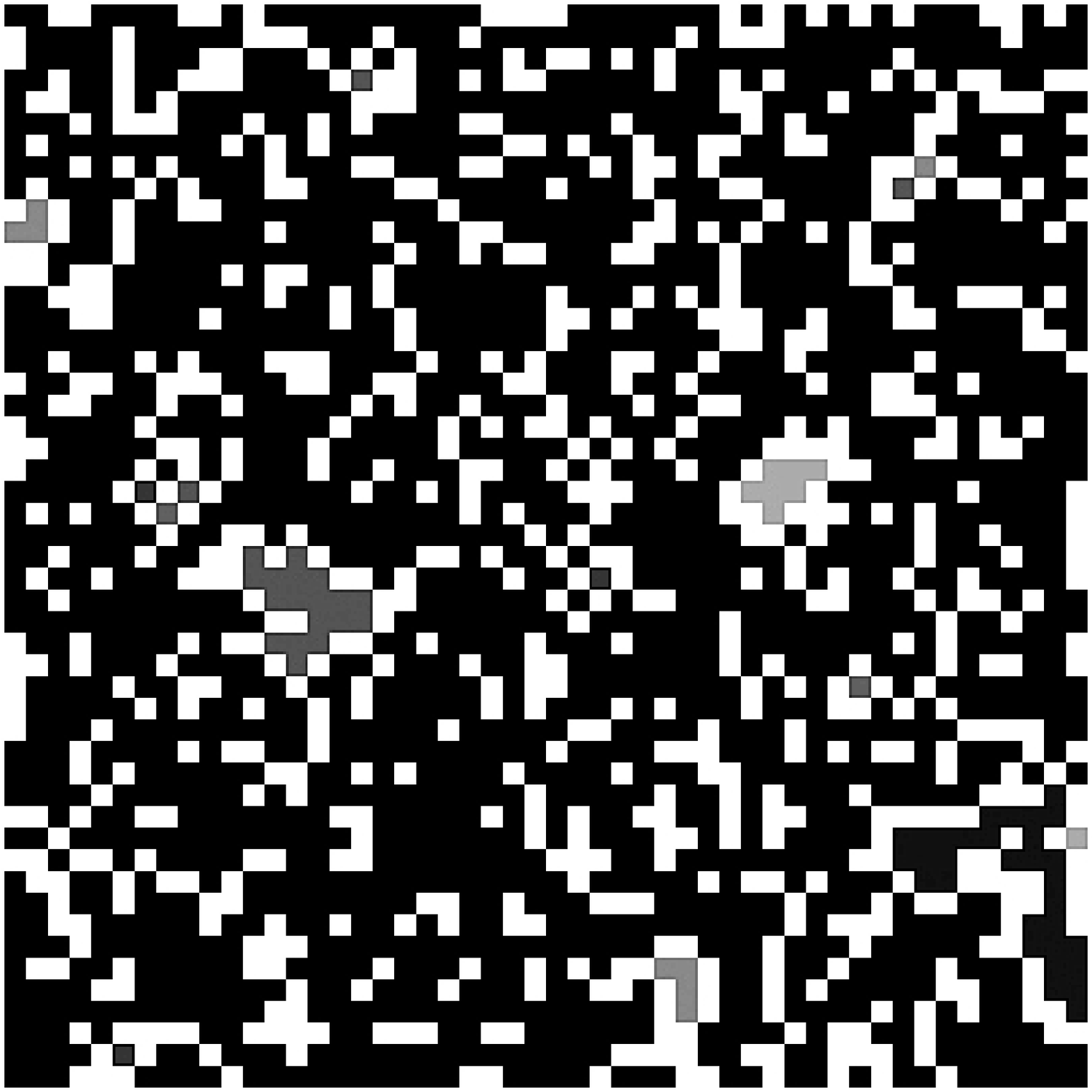,width=1.4in}~ \psfig{file=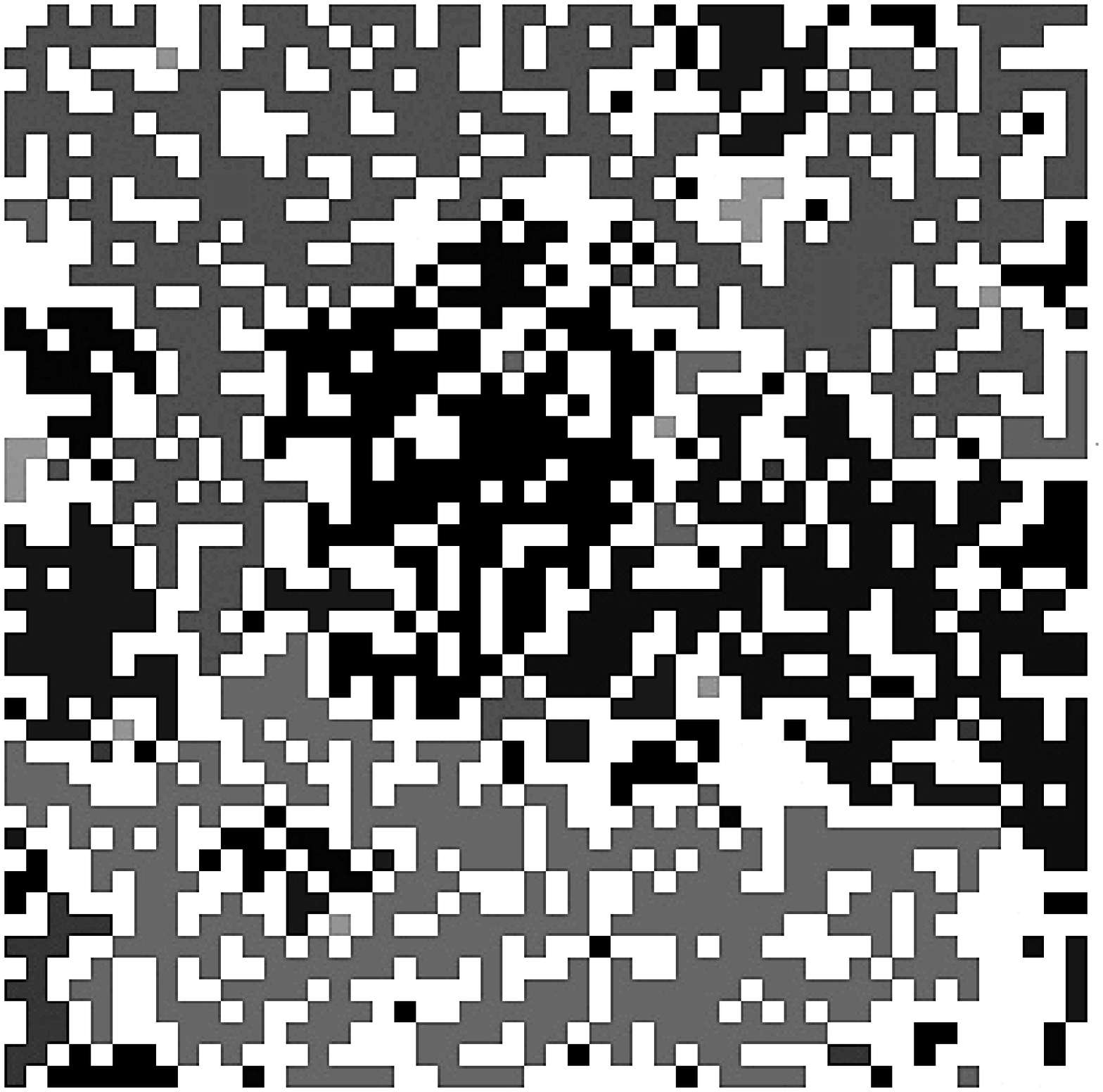,width=1.4in}~
\psfig{file=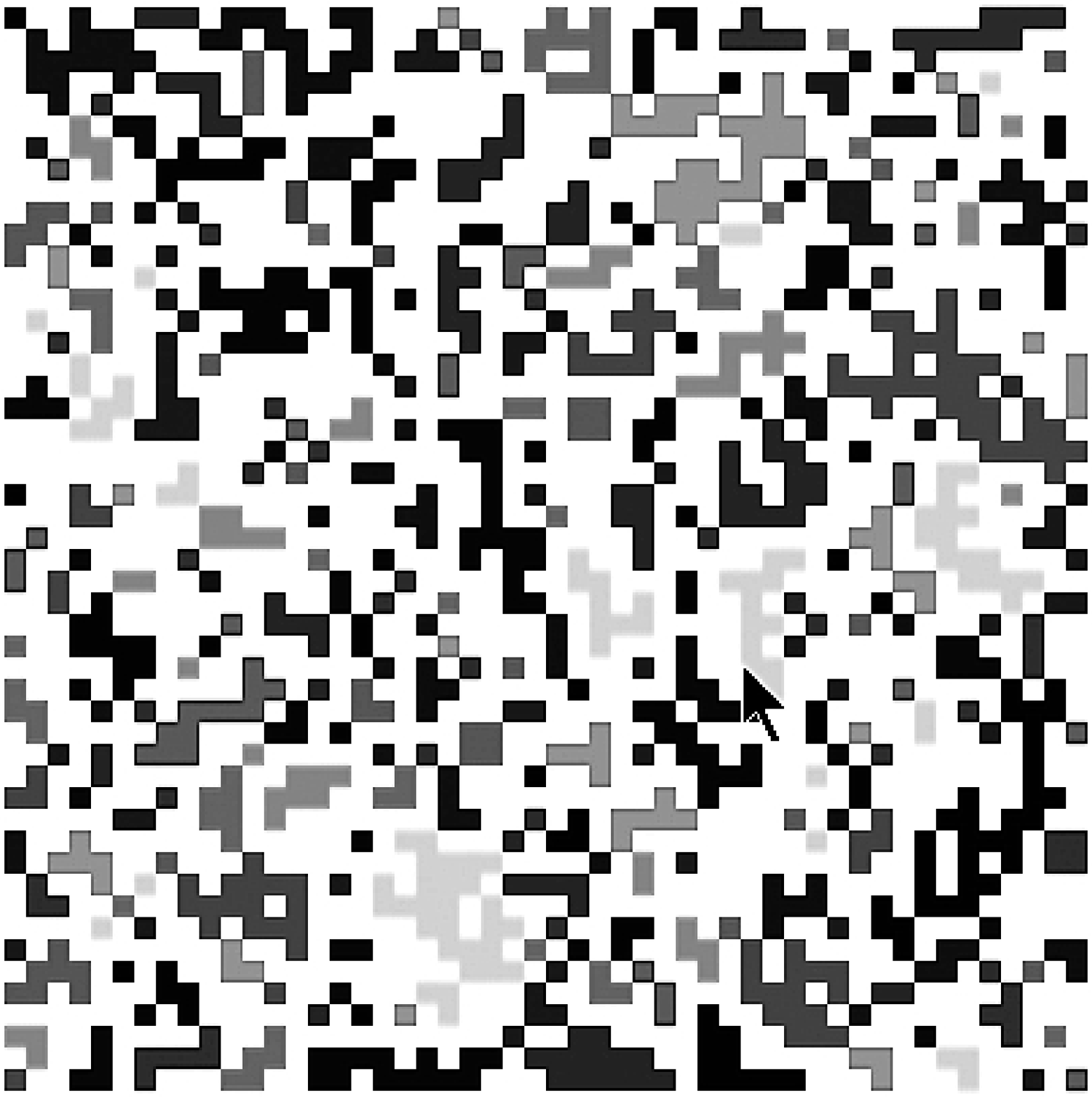,width=1.4in}}
\caption{Site-diluted square lattice at impurity concentrations below ($p=0.3$), at ($p=p_c\approx 0.4073$),
and above ($p=0.6$) the percolation threshold.}
\label{fig:percolation}
\end{figure}

Classical percolation theory\cite{StaufferAharony_book91} deals with the geometric
properties of randomly diluted lattices. The central question is whether the diluted
lattice contains a large cluster that spans the entire sample or whether it is decomposed
into small disconnected pieces. For definiteness, consider a $d$-dimensional hypercubic
lattice with bonds between the nearest neighbor sites in which a fraction $p$ of all
sites (site percolation) or bonds (bond percolation) is removed at random. Depending on
 $p$, the lattice can be in one of two ``phases'', separated by a sharp
percolation threshold at $p=p_c$. If $p<p_c$ (the percolating phase), there is a single
large cluster that spans the entire sample (as well as some smaller clusters). In the
thermodynamic limit, this cluster, the so-called \emph{infinite cluster}, becomes
infinitely large and contains a nonzero fraction $P_\infty$ of all sites. In contrast,
for $p>p_c$ the lattice is decomposed into small disconnected finite-size clusters only
(see Fig.\ \ref{fig:percolation}).
Right at $p=p_c$, there are clusters on all length scales, and their
structure is fractal.

The behavior of the diluted lattice close to $p_c$ is very similar to
the critical behavior near a continuous (2nd order) phase transition with the geometric
fluctuations due to dilution playing the role of the usual thermal or quantum
fluctuations. This implies that observables are governed by power-law scaling relations.
A central quantity is the cluster size distribution $n_s(p)$. It measures the number of
connected clusters with $s$ sites (per lattice site). Close to
$p_c$, it takes the scaling form
\begin{equation}
n_{s}(p) =s^{-\tau_c }f\left[ (p-p_c)\,s^{\sigma_c}\right]~,
\label{eq:cluster_size_distrib}
\end{equation}
where $\sigma_c$ and $\tau_c$ are critical exponents. In two dimensions, they are known
exactly, $\sigma_c=36/91$ and $\tau_c=187/91$; and in three dimensions they are well
known numerically, $\sigma_c \approx 0.45$ and $\tau_c \approx 2.18$. $f(x)$ is a scaling
function which behaves as $f(x) \sim \exp(-B_1 x^{1/\sigma_c})$ for $x>0$, $f(x) = {\rm
const}$ for $x=0$, and  $f(x) \sim \exp[{-(B_2 x^{1/\sigma_c})^{1-1/d}}]$ for $x<0$. The
critical behavior of all other geometric properties can be expressed in terms of the
exponents $\sigma_c$ and $\tau_c$. In particular, in the percolating phase, the fraction
of sites in the infinite cluster behaves as $P_\infty \sim (p_c -p)^{\beta_c}$ with the
exponent $\beta_c$ given by $\beta_c= (\tau_c-2)/\sigma_c$. When approaching the
percolation threshold, the typical linear size of the finite-size clusters, the
connectedness length, diverges as $\xi_c \sim |p-p_c|^{-\nu_c}$ with $\nu_c
=(\tau_c-1)/(d\sigma_c)$. Finally, the fractal dimension $D_f$ of the infinite cluster at
the percolation threshold can be expressed as $D_f = d/(\tau_c-1)$.

\section{Diluted classical magnets}\label{sec:classical}

\begin{figure}[t]
\centerline{\psfig{file=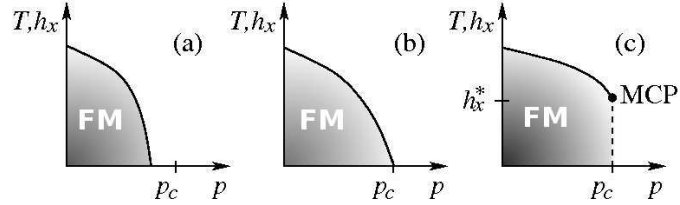,width=3.5in}} \caption{Schematic phase diagrams
for classical and quantum magnets on diluted lattices.
         In the classical case, magnetic order is destroyed by increasing the temperature $T$,
         in the quantum case by increasing the quantum fluctuations, e.g.,
         the transverse field $h_x$ in a quantum Ising magnet.}
\label{fig:pd_all}
\end{figure}

\begin{figure}[t]
\centerline{\psfig{file=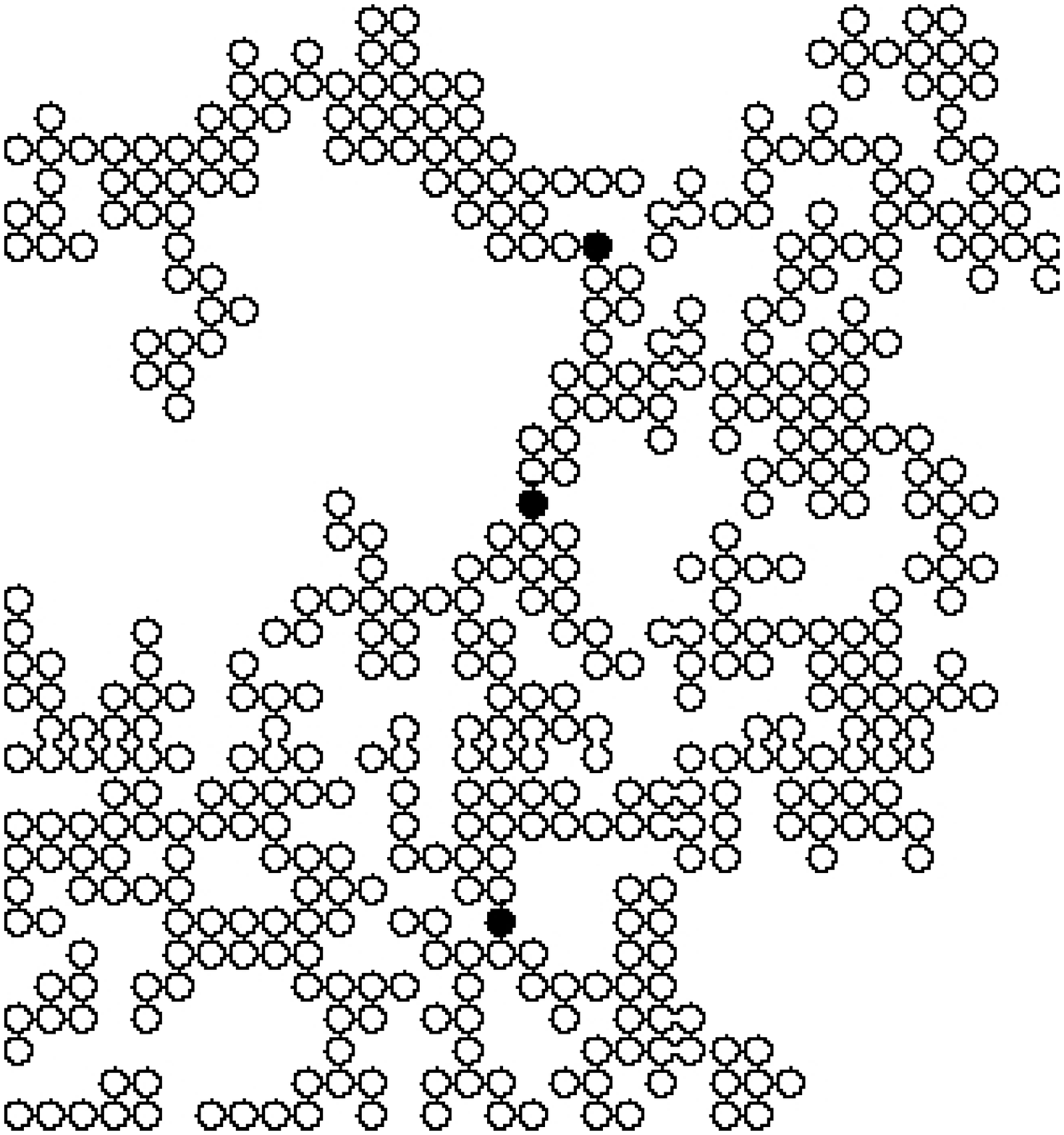,width=1.8in}
\psfig{file=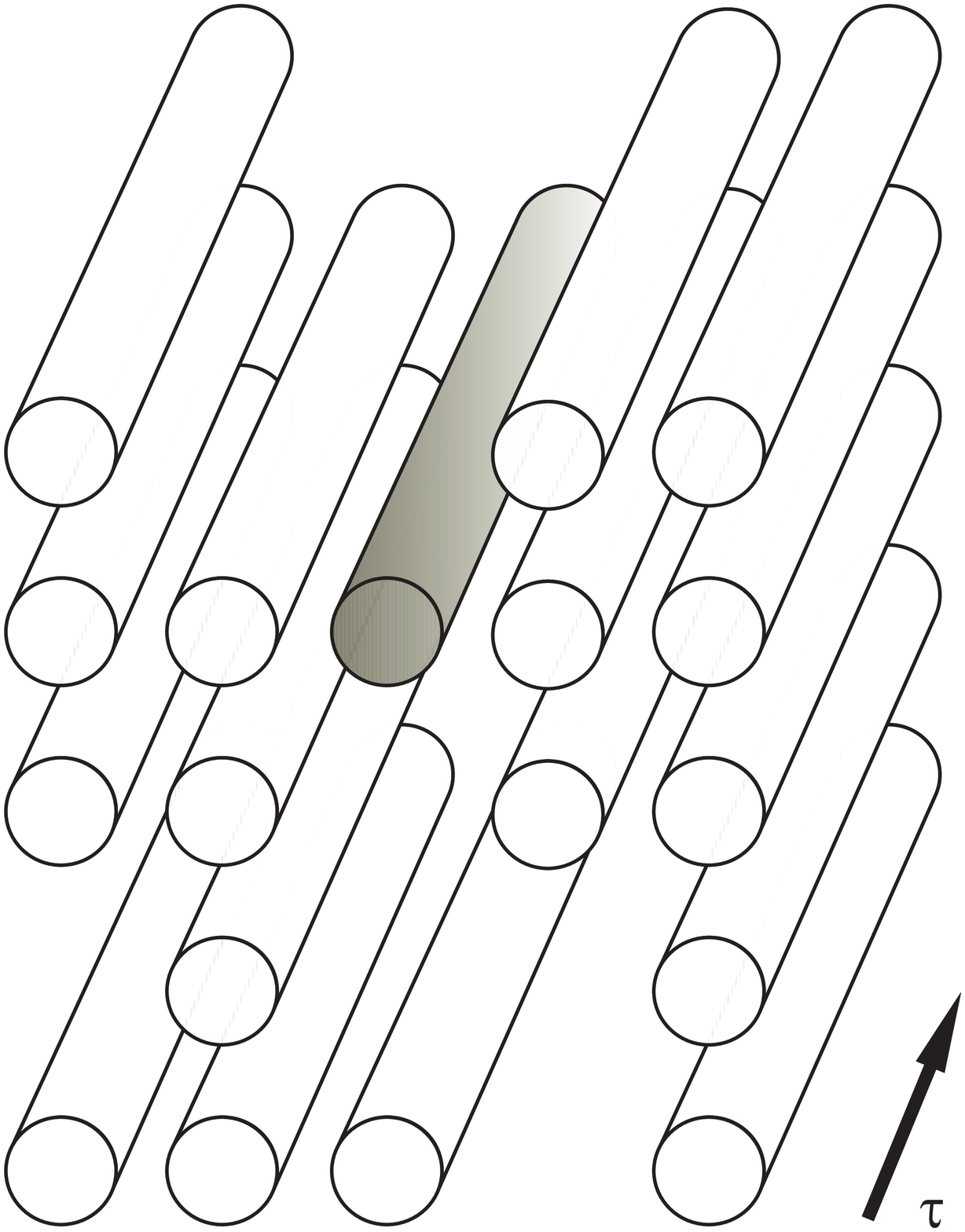,width=1.5in}} \caption{Red sites vs. red lines (shown in
black) in a critical percolation cluster.} \label{fig:red_sites}
\end{figure}

In this section we briefly summarize the behavior of a classical Ising
or Heisenberg magnet on a randomly diluted lattice. Consider the Hamiltonian
\begin{equation}
H = -J \sum_{\langle i,j\rangle} \epsilon_i \epsilon_j \, S_i S_j~,
\label{eq:classical_Ising}
\end{equation}
where $S_i$ is a classical Ising or Heisenberg spin at site $i$, and $J>0$
is the exchange interaction between nearest neighbors. The dilution is implemented
via quenched random variables $\epsilon_i$ taking the values
0 and 1 with probabilities $p$ and $1-p$, respectively. We first discuss the
temperature-dilution phase diagram. In the absence of dilution,
the model orders ferromagnetically below a critical temperature $T_c(0)$
(provided $d\ge 2$ for Ising and $d\ge 3$ for Heisenberg spins). Upon dilution, magnetic
order is weakened, and $T_c$ decreases. An important question is whether
the magnetic phase is completely destroyed before the dilution reaches $p_c$, right at
$p_c$, or whether long-range order survives even on the critical percolation cluster at
$p_c$, corresponding to phase diagrams (a), (b), and (c) in Fig.\ \ref{fig:pd_all},
respectively (long-range order cannot survive for $p>p_c$ because the system consists
of small disconnected clusters).
Phase diagram (a) can be excluded because the infinite percolation cluster is a massive
$d$-dimensional object for any $p<p_c$. Since the \emph{critical} percolation
cluster at $p=p_c$ has a fractal dimension $D_f>1$, one might be
tempted to conclude that it supports magnetic long-range order, at least in the Ising
case (implying a phase diagram of type (c)). However, this is incorrect: At the
percolation threshold, thermal fluctuations immediately destroy the magnetic
order.\cite{Bergstresser77,StephenGrest77,GefenMandelbrotAharony80}
This can be understood by considering the ``red sites'' of the critical percolation
cluster, i.e., sites that divide the cluster into two otherwise disconnected pieces
(see Fig.\ \ref{fig:red_sites}).
The orientation of the spins on these two pieces can be flipped with respect to each other
with a finite energy cost of $2J$. For any $T\ne 0$, both the parallel and
antiparallel configuration at each of the red sites contribute to the statistical sum,
destroying magnetic long-range order. The phase diagram is thus of type (b), and a
percolation transition only occurs at exactly
zero temperature. Since there are no thermal fluctuations, the critical behavior of this
transition is identical to geometric percolation. Any $T \ne 0$ destroys the
percolation critical behavior, instead the transition is in the universality class of the
corresponding generic disordered classical magnet.

\section{Diluted quantum magnets}\label{sec:quantum}

We now turn to the main topic, QPTs
on percolating lattices. Generally, QPTs occur at zero
temperature as functions of pressure, magnetic field
or other nonthermal control parameters (for reviews, see, e.g.,  Refs.\
\refcite{SGCS97,Sachdev_book99,Vojta_review00,BelitzKirkpatrickVojta05}).
One important aspect of these transitions is the so-called
quantum-to-classical mapping. It arises because in quantum statistical mechanics
the partition function does not factorize in potential and kinetic parts.
Instead, it has to be formulated in terms of space and time-dependent
variables. As a result, (imaginary) time acts like an additional coordinate, and
a QPT in $d$ dimensions can be related to a
classical transition in a higher dimension,\footnote{This mapping
is restricted to the thermodynamics only. Moreover some quantum transitions
lead to extra complications such as Berry phases that do not have a classical
counterpart.} a fact we will be using repeatedly below.

\subsection{Transverse-field Ising model}\label{subsec:ising}

The first example is a randomly diluted Ising model in a transverse magnetic field,
given by the Hamiltonian
\begin{equation}
\hat H_I = -J \sum_{\langle i,j\rangle} \epsilon_i \epsilon_j \, \hat S_i^z \hat S_j^z -
         h_x \sum_i \epsilon_i \hat S_i^x - H \sum_i \epsilon_i \hat S_i^z~.
\label{eq:quantum_Ising}
\end{equation}
$\hat S_i^z$ and  $\hat S_i^x$ are the $z$ and $x$ components of the the quantum spin-1/2
at site $i$; $h_x$ is the transverse field that controls the quantum fluctuations, and
$H$ is the field conjugate to the order parameter. The clean model ($p=0$) is a paradigm
for the study of QPTs: For $h_x \ll J$, the ground state is ferromagnetically ordered in
$z$-direction while for $h_x \gg J$ the quantum fluctuations due to the transverse field
destroy the long-range order. The two phases are separated by a QPT at $h_x \sim
J$.\cite{Sachdev_book99}

As in the classical case, we ask how the dilution influences the phase diagram (in the
$h_x$-$p$ plane), i.e., is the phase diagram of type (a), (b), or (c) in Fig.\
\ref{fig:pd_all}? As before, (a) can be excluded because the infinite percolation cluster
is a massive $d$-dimensional object for $p<p_c$. To decide between (b) and (c), we adopt
the ``red-site'' argument to the quantum case. Following the quantum-to-classical
mapping, we have to consider an effective system in $d$ space dimensions and one extra
imaginary time dimension which becomes infinite for temperature $T\to 0$. Crucially, the
defect positions are time-independent. Instead of red sites we thus have ``red lines''
separating different pieces of the critical percolation cluster (right panel of Fig.\
\ref{fig:red_sites}). Configurations with different spin orientations on two such pieces
now come with a infinite effective energy penalty and are suppressed. This suggests that
magnetic long-range order can survive on the critical percolation cluster if the quantum
fluctuations are not too strong, implying a phase diagram of type (c). This has been
confirmed by simulation results not only for quantum Ising models but also Heisenberg
magnets and quantum rotors.\cite{Harris74b,Sandvik02,VojtaSknepnek06}

\begin{figure}[t]
\centerline{\psfig{file=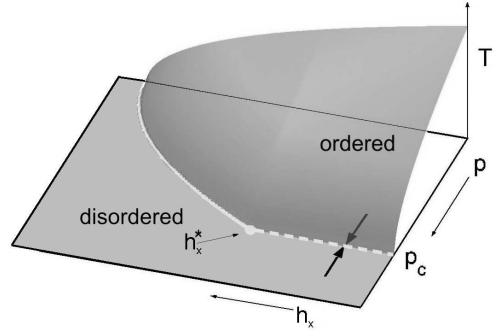,width=2.6in}}
\caption{Schematic $T-h_x-p$ phase diagram of a diluted quantum magnet. There is
    a multi-critical point at $(T=0, h_x^\ast,p_c)$. The QPT
    across the dashed line is the topic of this paper.}
\label{fig:3d_pd}
\end{figure}

Combining the effects of thermal and quantum fluctuations, we obtain the $T-h_x-p$
phase digram shown in Fig.\ \ref{fig:3d_pd}.
A diluted quantum magnet can thus undergo two nontrivial QPTs, separated by a
multicritical point at $(T=0, h_x^\ast,p_c)$. We are interested in the percolation
transition at $p_c$ and $h_x<h_x^\ast$, i.e., the transition across the dashed line in
Fig.\ \ref{fig:3d_pd}. It was first investigated in detail by Senthil and Sachdev
\cite{SenthilSachdev96}: Consider a single percolation cluster of $s$ sites. For small
transverse field $h_x<h_x^\ast$, all spins on the cluster are parallel. The cluster thus
acts as a two-level system with an energy gap (inverse susceptibility) $\Delta_s$ that
depends exponentially on the cluster size $\Delta_s \sim \chi_s^{-1} \sim h_x \exp(-Bs)$
with $B \sim \ln(J/h_x)$. (All other excitations have at least energy $J$.) Since the
cluster size $s$ and its linear extension $L$ are related via the fractal dimension $s
\sim L^{D_f}$, we obtain an unusual exponential relation between length and (inverse)
time scales which is sometimes called activated scaling,
\begin{equation}
\ln (h_x/\Delta_s) \sim L^{D_f}~. \label{eq:activated}
\end{equation}

The critical behavior of the total system can now be found by summing over all
percolation clusters via the cluster size distribution (\ref{eq:cluster_size_distrib}).
Let us first consider static quantities like magnetization or magnetic spatial
correlation length. For $h_x<h_x^\ast$, magnetic long-range order survives on the
infinite percolation cluster, while all finite-size clusters do not contribute. Thus, the
total magnetization is proportional to the number of sites in the infinite cluster, $m
\sim P_\infty \sim (p_c-p)^{\beta_c}$ for $p<p_c$. The magnetic order parameter exponent
$\beta$ is thus identical to that of geometric percolation. A similar argument can be
made for the magnetic correlation length $\xi$: For $h_x<h_x^\ast$, all spins on a
cluster are correlated, but the correlations cannot extend beyond the cluster size, thus
$\xi \sim \xi_c \sim |p-p_c|^{-\nu_c}$, and the magnetic correlation length exponent is
identical to the geometric one, too.
In contrast, quantities involving quantum dynamics behave nonclassically. For instance,
the dependence of the magnetization on the ordering field $H$ takes the scaling form
\begin{equation}
m(p-p_c,H) = b^{-\beta_c/\nu_c}~ m\left( (p-p_c)b^{1/\nu_c}, \ln(H)b^{-D_f} \right)~,
\end{equation}
with $b$ being an arbitrary scale factor. At the percolation threshold $p=p_c$,
this gives the unconventional relation $m \sim [\ln(H)]^{2-\tau_c}$.
For $p\ne p_c$, the transition is accompanied by strong power-law quantum
Griffiths effects.\cite{Vojta06,SenthilSachdev96}

\subsection{Bilayer quantum Heisenberg magnet}\label{subsec:bilayer}

\begin{figure}[t]
\centerline{\psfig{file=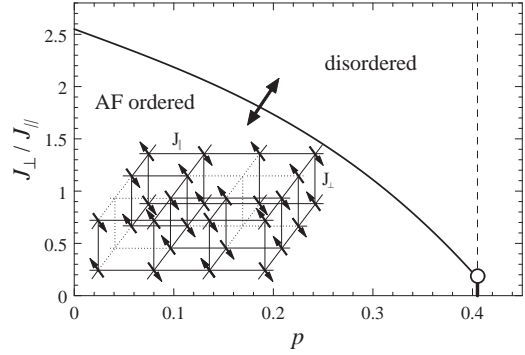,width=2.8in}}
\caption{Phase diagram of the dimer-diluted bilayer Heisenberg antiferromagnet
         (after Refs.\ \refcite{Sandvik02,VajkGreven02}).
         Inset: Sketch of the system.}
\label{fig:bilayer}
\end{figure}

This subsection is devoted to diluted Heisenberg magnets. Specifically, we consider a
dimer-diluted bilayer quantum Heisenberg antiferromagnet with the Hamiltonian
\begin{equation}
\hat H_H=J_\parallel \sum_{{\langle i,j\rangle} \atop a=1,2}\epsilon_i\epsilon_j
\mathbf{\hat{S}}_{i,a}\cdot{\mathbf{\hat{S}}}_{j,a}+ J_\perp\sum_i
\epsilon_i{\mathbf{\hat{S}}}_{i,1}\cdot{\mathbf{\hat{S}}}_{i,2}~, \label{eq:bilayer_H}
\end{equation}
where $\hat S_{j,a}$ is the spin operator at site $j$ in layer $a=1$ or 2. The clean
system ($p=0$) undergoes a QPT between a paramagnetic and an antiferromagnetic phase as a
function of the ratio $J_\perp/J_\parallel$ between the inter-layer coupling and the
in-plane interaction.
 For $J_\perp \gg J_\parallel$,
the corresponding spins in the two layers form a singlet which is magnetically inert.
Thus, there is no long-range order. In contrast, for $J_\parallel \gg J_\perp$, each
layer orders antiferromagnetically, and the weak inter-layer coupling establishes
antiferromagnetic order between the layers. The phase diagram of the dimer-diluted system
has been determined by Sandvik\cite{Sandvik02} and Vajk and Greven\cite{VajkGreven02}; it
is shown in Fig.\ \ref{fig:bilayer}.
In agreement with the general arguments given in the last subsection,
long-range order can survive at $p=p_c$ giving rise to a nontrivial
percolation QPT (across the short vertical line
in Fig.\ \ref{fig:bilayer}). To study this transition, we first map
the low-energy physics of (\ref{eq:bilayer_H}) onto a quantum rotor
model with the action\cite{Sachdev_book99}
\begin{equation}
\mathcal{A} =\int d\tau \sum_{\langle ij\rangle }J_\parallel \epsilon _{i}\epsilon
_{j}\mathbf{S}_{i}(\tau )\cdot \mathbf{S}_{j}(\tau )+ \frac{T}{g}\sum_{i}\sum_n \epsilon _{i} \,\omega
_{n}^{2}\, \mathbf{S}_i(\omega _{n})\mathbf{S}_i(-\omega _{n})~.
\label{eq:NLSM}
\end{equation}
Here, each rotor variable $\mathbf{S}_{i}(\tau )$ (a unit vector at site $i$ and imaginary time $\tau$),
describes a dimer of corresponding spins in the two layers.
$\omega_n$ is a Matsubara frequency, and the parameter $g$ is related to the ratio
$J_\perp/J_\parallel$ of the quantum Hamiltonian (\ref{eq:bilayer_H}).

Our approach\cite{VojtaSchmalian05b} is the same as in Sec.\ \ref{subsec:ising}, we first
consider a single percolation cluster of size $s$ and then sum over all clusters by means
of the cluster size distribution  (\ref{eq:cluster_size_distrib}). For small $g$, all
rotors on the cluster are correlated but collectively fluctuate in time. Thus, each
cluster acts as a (0+1)-dimensional rotor model with magnetic moment $s$. Its low-energy
properties can be easily found by a renormalization group calculation or dimensional
analysis, leading to a scaling form of the free energy
\begin{equation}
F_{s}\left( g,H,T\right) =(g/s) \Phi \left( Hs^{2}/g,Ts/g\right)
\label{eq:F_s}
\end{equation}
as a function of $g$, $T$, and magnetic field $H$. Here, $\Phi$ is a universal scaling
function. Eq.\ (\ref{eq:F_s}) implies that the thermodynamics of a quantum spin cluster
is more singular in its size $s$ than that of a classical cluster. In particular,
classically, the magnetic susceptibility increases like $\chi^c_s \sim s^2$ while in our
quantum model at $T=0$, it increases more strongly, $\chi_s \sim s^{3}$. The dynamical
critical exponent can be obtained by relating the gap $\Delta$ to the linear size $L$ of
the cluster via $\chi_s \sim s^{2}/\Delta$ giving $\Delta \sim s^{-1} \sim L^{-D_f}$.
Thus, the dynamical exponent at the percolation QPT is $z=D_f$. The total free energy is
obtained by summing (\ref{eq:F_s}) over all clusters. This gives rise to the general
scaling scenario
\begin{eqnarray}
2-\alpha  &=&\left( d+{z}\right) \nu~,  \\
 \beta  &=&\left( d-D_{f}\right) \nu~,  \\
 \gamma &=&\left( 2D_{f}-d+{z}\right) \nu~,  \\
 \delta  &=&({D_{f}+{z}})/({d-D_{f}})~, \\
2-\eta &=&  2D_{f}-d+{z}~.
\label{exponents}
\end{eqnarray}
All critical exponents are completely determined by two geometric percolation exponents
(say $D_f$ and $\nu=\nu_c$) together with the dynamical exponent $z$ which contains the information on the quantum
fluctuations. Thus, $\alpha $, $\gamma $, $\delta $, and $\eta$ are modified by the
quantum dynamics  while $\beta$ and $\nu$ are unchanged. The resulting exponent values
are shown in table \ref{tab:exponents}.
\begin{table}[tbp]
\tbl{Critical exponents of the geometric and quantum percolation transition in
         two and three dimensions.\cite{VojtaSchmalian05b}}
{\begin{tabular}{c|cc|cc}
\hline & \multicolumn{2}{c}{2d} & \multicolumn{2}{c}{3d} \\
 exponent      & classical & quantum & classical & quantum \\
\hline
$\alpha $ & $-$2/3 & $-$115/36 & $-$0.62 & $-$2.83 \\
$\beta $ & 5/36 & 5/36 & 0.417 & 0.417 \\
$\gamma $ & 43/18 & 59/12 & 1.79 & 4.02 \\
$\delta $ & 91/5 & 182/5 & 5.38 & 10.76 \\
$\nu $ & 4/3 & 4/3 & 0.875 & 0.875 \\
$\eta $ & 5/24 & $-$27/16 & $-$0.06 & $-$2.59 \\
$z$ & - & 91/48 & - & 2.53 \\ \hline
\end{tabular}}
\label{tab:exponents}
\end{table}
We have recently confirmed the 2d results by performing large-scale Monte-Carlo simulations
of the action (\ref{eq:NLSM}).\cite{SknepnekVojtaVojta04,VojtaSknepnek06} Let us note
that the behavior of site-diluted (rather than dimer-diluted) Heisenberg antiferromagnets is
more complicated because the effective action contains Berry phases. Recent computer
simulations\cite{WangSandvik06} suggest that $z \approx 1.5 D_f$ to 2$D_f$ in this case.

\subsection{Percolation and dissipation}\label{subsec:dissipation}

In many real systems, magnetic degrees of freedom are coupled to a dissipative environment of
``heat bath modes'' (e.g., electronic degrees of freedom in a metal or nuclear spins
in molecular magnet). In this subsection, we study the influence of
Ohmic dissipation on a percolation QPT.\cite{HoyosVojta06}
To this end we add
baths of harmonic oscillators to the diluted transverse-field Ising model
of Sec.\ \ref{subsec:ising},
\begin{equation}
\hat H = \hat H_I + \sum_{i,n} {\epsilon_i} \left[{\nu_{i,n}}a_{i,n}^{\dagger}a_{i,n}+
\frac 1 2 {\lambda_{i,n}}\hat S_{i}^{z} (a_{i,n}^{\dagger}+a_{i,n}) \right]~,
\label{eq:dissipative_H}
\end{equation}
where $a_{i,n}$ and $a_{i,n}^{\dagger}$ are the annihilation and creation operators of
the $n$-th oscillator coupled to spin $i$; $\nu_{i,n}$ is its frequency, and
$\lambda_{i,n}$ is the coupling constant. All baths have the same spectral function
${\cal E}(\omega)=\pi \sum_{n}\lambda_{i,n}^{2} \delta (\omega-\nu_{i,n})/\nu_{i,n}=2\pi
\alpha\omega e^{-\omega/\omega_{c}}$ with $\alpha$ the dimensionless dissipation strength
and $\omega_{c}$ the cutoff energy.

\begin{figure}[t]
\centerline{\psfig{file=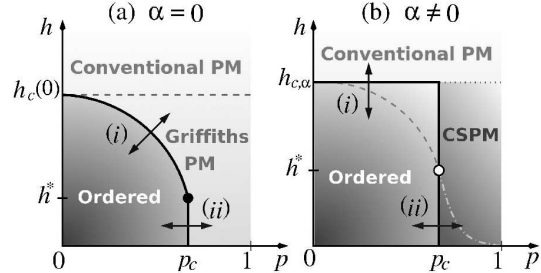,width=2.8in}}
\caption{Schematic ground state phase diagrams of the diluted transverse-field Ising magnet without
(a) and with (b) dissipation. CSPM is the classical superparamagnetic phase (after Ref.\ \refcite{HoyosVojta06}).}
\label{fig:dissipative_pd}
\end{figure}

Following our general approach we first consider a single percolation cluster of size
$s$. Without dissipation, its low-energy properties are described by a quantum-mechanical
two-level system (see Sec.\ \ref{subsec:ising}). In the presence of the heat baths, the
cluster therefore behaves as a dissipative two-level system with effective dissipation
strength $s\alpha$. The physics of this problem is very rich, it has been reviewed, e.g.,
in Ref.\ \refcite{LCDFGZ87}. For our purposes, the most important aspect is that with
increasing dissipation strength, the (Ohmic) dissipative two-level system undergoes a
phase transition from a fluctuating phase at $s\alpha<1$ to a localized (frozen) phase at
$s\alpha>1$. As a result, for any given microscopic dissipation strength $\alpha$, the
total diluted lattice consists of a mixture of large frozen clusters that act as
classical superspins and smaller clusters that behave quantum mechanically down to the
lowest temperatures. The resulting phase diagram\cite{HoyosVojta06} of the of the
dissipative diluted transverse-field Ising model is shown in Fig.\
\ref{fig:dissipative_pd}.

The behavior of observables close to the percolation transition can be obtained by
summing the results for the dissipative two-level system over the cluster size
distribution (\ref{eq:cluster_size_distrib}). The total magnetization has 3 parts:
The infinite percolation cluster, if any, contributes
$m_\infty \sim P_\infty \sim (p_c-p)^{\beta_c}$. The frozen finite size clusters
individually have nonzero magnetization, but they do not align in the absence of an
ordering field. Finally, the small fluctuating clusters have vanishing magnetization.
The interplay between these 3 contributions and an ordering field leads to exotic
hysteresis phenomena.\cite{HoyosVojta06} The low-temperature susceptibility is dominated
by the frozen clusters and behaves classically, $\chi \sim |p-p_c|^{\gamma_c}/T$.
In contrast, the specific heat is determined by quantum fluctuations giving
$C \sim 1/\ln^{2}(h_x/T)$.

\section{Conclusions}\label{sec:conclusions}

In summary, we have discussed the interplay of geometric, thermal and quantum
fluctuations at the percolation threshold. While thermal fluctuations immediately
destroy magnetic long-range order on the critical percolation cluster,
the effects of quantum fluctuations are more subtle. Generically, magnetic long-range
order on the critical percolation cluster can survive a finite amount of quantum
fluctuations. This gives rise to a nontrivial percolation QPT and
a multicritical point separating it from the generic ``disordered'' transition
at $p<p_c$.

We have discussed three examples of such percolation QPTs in quantum Ising and Heisenberg
magnets with and without dissipation. In all cases, the critical behavior is different
from classical (geometric) percolation, but it can be expressed in terms of the geometric
percolation critical exponents. Percolation transitions are thus among the very few
examples of QPTs with exactly known exponent values in two dimensions. This is caused by
the fact that at our percolation transitions, the criticality is due to the
\emph{geometric criticality} of the underlying diluted lattice. However, the quantum
fluctuations ``go along for the ride'' and modify the behavior of all quantities
involving dynamic correlations.

In the quantum Heisenberg case (Sec.\ \ref{subsec:bilayer}), this leads to new critical exponents while the
overall power-law scaling scenario is still valid. In contrast, for the transverse-field
Ising model (Sec.\ \ref{subsec:ising}), the dynamical scaling is of activated
(exponential) rather than the usual power-law type. Finally, in the presence of
dissipation (Sec.\ \ref{subsec:dissipation}), the singularities in the dynamics become
even stronger such that individual finite-size clusters can undergo the phase transition
independently from the bulk. This leads to a novel superparamagnetic classical cluster
phase. Note that these different cases agree with a general classification
of phase transitions in the presence of disorder\cite{Vojta06} based on the effective
dimensionality of the defects.

\section*{Acknowledgements}

We gratefully acknowledge discussions with M.\ Greven, S. Haas, H.\ Rieger,
A.\ Sandvik, J.\ Schmalian, and M.\ Vojta. Parts of this work have been performed
at the Aspen Center for Physics and the Kavli Institute for Theoretical Physics,
Santa Barbara. This work has been supported by the NSF under grant no. DMR-0339147,
by Research Corporation, and by the University of Missouri Research Board.

\bibliographystyle{ws-procs975x65}
\bibliography{../00bibtex/rareregions}

\end{document}